\RequirePackage{fix-cm}
\documentclass[smallextended]{svjour3}       % onecolumn (second format)
\smartqed  % flush right qed marks, e.g. at end of proof
\usepackage{graphicx}
\usepackage{epstopdf}
\begin{document}

\title{Microscopic laws vs.~Macroscopic laws: Perspectives from kinetic theory and hydrodynamics%\thanks{Grants or other notes
%about the article that should go on the front page should be
%placed here. General acknowledgments should be placed at the end of the article.}
}
%\subtitle{Do you have a subtitle?\\ If so, write it here}

%\titlerunning{Short form of title}        % if too long for running head

\author{Mahendra K. Verma }

%\authorrunning{Short form of author list} % if too long for running head

\institute{Mahendra K. Verma \at
             Department of Physics, Indian Institute of Technology Kanpur, Kanpur 208016, India \\
              Tel.: +91-5122597396\\
              \email{mkv@iitk.ac.in}           %  \\
%             \emph{Present address:} of F. Author  %  if needed
}

\date{Received: date / Accepted: date}
% The correct dates will be entered by the editor

\maketitle

\begin{abstract}
{\em Reductionism} is a prevalent viewpoint in science according to which all physical phenomena can be understood from fundamental laws of physics.  Anderson~[Science, 177, 393 (1972)], Laughlin and Pines~[PNAS, 97, 28 (2000)], and others have countered this viewpoint and argued in favour hierarchical structure of the universe and laws.  In this paper we advance the latter perspective by showing that some of the complex flow properties derived using hydrodynamic equations (macroscopic laws) are very difficult, if not impossible, to describe in microscopic framework---kinetic theory.  These properties include Kolmogorov's theory of turbulence, turbulence dissipation and diffusion, and dynamic pressure.  We also provide several other examples of hierarchical description. 
\keywords{Hydrodynamics \and Kinetic theory \and Hierarchical laws }
% \PACS{PACS code1 \and PACS code2 \and more}
% \subclass{MSC code1 \and MSC code2 \and more}
\end{abstract}

\section{Introduction}
\label{intro}
A prevalent view in science is that all phenomena in the universe can ``in principle'' be explained using fundamental laws of physics.  This paradigm, called {\em reductionist hypothesis}, encouraged search for microscopic laws that led to fascinating discoveries in quantum mechanics and particle physics~\cite{Kane:book:Particle}.    Buoyed by the success of these discoveries, some  physicists are looking for a reductionist framework that can explain all the physical phenomena of the universe.  This holy grail is referred to as  {\em theory of everything (TOE),  final theory, ultimate theory}, and {\em master theory}~\cite{Weinberg:book:dream,Hawking:book:TOE}.  The aforementioned viewpoint has many champions and supporters, but it has also invited criticisms as descried below.

The degree of criticism and support to the reductionist paradigm vary.  For example, Weinberg~\cite{Weinberg:book:dream} argues strongly in favour of reductionism, and claims that all scientists, including economists, practice reductionism.  According to Weinberg, ``it saves scientists from wasting their ideas that are not worth pursuing'', and/or provides stronger theoretical basis for their hypothesis.  Refer to \cite{Weinberg:book:dream,Hawking:book:TOE} for more references  in support of reductionism.

In a somewhat sharp criticism,  Anderson~\cite{Anderson:Science1972} argued that ``the reductionism hypothesis does not by any means imply a `constructionist' one: the ability to reduce everything to simple fundamental laws does not imply the ability to start from those laws and reconstruct the universe''.  Further he agues that if the starting point of a field Y is field X, then it does not mean that all the laws of Y are ``just applied X''.  He goes on the illustrate the above viewpoint by showing how the ideas of broken symmetries  (apart from fundamental laws) help explain diverse phenomena of condensed matter physics. 

In another article critical of reductionism,  Laughlin and Pines~\cite{Laughlin:PNAS2000} write ``The emergent physical phenomena regulated by higher organizing principles have a property, namely their insensitivity to microscopics, that is directly relevant to the broad question of what is knowable in the deepest sense of the term.''  They further argue, ``Rather than a Theory of Everything we appear to face a hierarchy of Theories of Things, each emerging from its parent and evolving into its children as the energy scale is lowered.'' Also refer to Laughlin~\cite{Anderson:book:More,Laughlin:book}.

Another set of illustrations on limitations of reductionism are as follows.  The letters of the book do not convey the story of a book.  A combination of words, paragraphs, and chapters that describe subplots and plots makes the story.  Similarly, music and paintings cannot be appreciated by just focussing on musical notes and photon packets; rather, they are complex hierarchical structures with notes and colours appearing at the bottom-most layer.  The aesthetics and ecology of a building is impossible to derive from the properties of bricks and mortar.  A complex computer programs is a hierarchical structure with program statements, functions, data structures, and their combinations  (called {\em classes});  it is very difficult to decipher the  functionality of a program if we focus only on the program statements.   Carrying the analogy to physics, though every macroscopic physical system is made of electrons and protons, its macroscopic properties  follow from the complex organization of different things. For the Earth, we need to focus on the macroscopic objects like atmosphere, oceans, lakes, land, life, etc., rather than electrons and protons that make them.
 
 After so many discussion by eminent scientists, it appears futile to write more on this topic.  However in the present article, I  provide several interesting examples of hydrodynamic laws (a macroscopic description) that  cannot be conveniently derived using the microscopic counterpart, for example, kinetic theory. These examples provide much simpler comparison between microscopic and macroscopic laws, in comparison to more complex ones involving stars, planets, biology, society, etc.     The present article  essentially advances the viewpoint that not all macroscopic phenomena can be explained from microscopic perspectives  \cite{Anderson:Science1972,Laughlin:PNAS2000}.

\section{Kinetic theory and hydrodynamics}
\label{sec:KT}
 In kinetic theory, we deal with a large number of particles (say $N$) that are specified by their position (${\bf r}$) and velocity (${\bf u}$).  These particles are represented as a point in $6N$-dimensional phase space whose coordinates are $(x_a, y_a, z_a, p_{x,a}, p_{y,a}, p_{z,a})$, where $a$ is the particle label; or as $N$ points in a six-dimensional $\mu$-space whose coordinates are $(x, y, z, p_x, p_y, p_z)$.    The density of these points in $\mu$-space is called {\em distribution function}, and it is denoted by $f({\bf r,u}, t)$~\cite{Choudhuri:book:Fluids}.  The Boltzmann equation of kinetic theory describes the evolution of the distribution function, and it is the starting point for many works of statistical physics~\cite{Lifshitz:book:Physical_Kinetics,Choudhuri:book:Fluids,Liboff:book}.  Kinetic theory successfully describes  many phenomena---thermodynamics; phase transitions;  observed properties of gas, liquids, polymers; etc.

 On the other hand, hydrodynamic description involves real-space density $\rho({\bf r})$, velocity ${\bf u(r)}$, and internal energy $e({\bf r})$~\cite{Landau:book:Fluid}.  The equations of these variables were derived in continuum framework by Euler, Navier, Stokes, and others.  These equations are essentially Newton's laws of motion for fluid elements in the flow.   Here, the field variables are averaged quantity over many microscopic particles.  This is called {\em continuum approximation}.    Note however that hydrodynamic equations can be derived using using kinetic theory.   An averaging of the Boltzmann equation (with collision terms) and its various moments  yields equations for $\rho({\bf r})$,  ${\bf u(r)}$, and $e({\bf r})$~\cite{Lifshitz:book:Physical_Kinetics,Choudhuri:book:Fluids,Liboff:book}.  Such derivations are popular among the astro- and plasma physicists.  
 
In the following discussion we will describe several important hydrodynamic laws---Kolmogorov's theory of turbulence, irreversibility in turbulence, accelerated diffusion in turbulence, dynamic pressure, etc., which could be treated as macroscopic laws since they are derived using a multiscale description of hydrodynamic equations.    We show in the next several sections that the above laws  cannot be   derived conveniently starting solely from  kinetic theory.   As far as we know, no one provided such derivations from the first principles.   Note that even derivation of incompressible hydrodynamics from kinetic theory itself is quite difficult~\cite{Bisi:JPA2014}.

%%%
\section{Multiscale energy transfers and flux in hydrodynamic turbulence}
Many natural (astrophysical and geophysical) and engineering flows are turbulent.  Generic features among them are---energy feed at the large scales, and energy flow to smaller and smaller scales that finally gets converted to heat.  See Fig.~1 for an illustration.  This multiscale feature has been propounded by Richardson, Taylor, Prandtl, Kolmogorov, and others~\cite{Kolmogorov:DANS1941Dissipation,Kolmogorov:DANS1941Structure,Frisch:book,Pope:book,Lesieur:book:Turbulence,McComb:book:Turbulence}.  According to Kolmogorov, in incompressible hydrodynamic turbulence forced at large scales, the energy flux at the intermediate scale is constant ($\epsilon_u$), while the velocity fluctuations $u_l \sim (\epsilon_u l)^{1/3}$. The corresponding energy spectrum is $E_u(k) = K_\mathrm{Ko} \epsilon_u^{2/3} k^{-5/3}$, where $K_\mathrm{Ko}$ is Kolmogorov's constant, and $k$ is wavenumber.  The  multiscale energy transfer of Fig.~1 has been derived both in real space and Fourier space formulation of hydrodynamic turbulence~\cite{Kolmogorov:DANS1941Dissipation,Kolmogorov:DANS1941Structure,Frisch:book,Pope:book,Lesieur:book:Turbulence,McComb:book:Turbulence}.

\begin{figure}%[tbhp]
\centering
\includegraphics[width=0.8\linewidth]{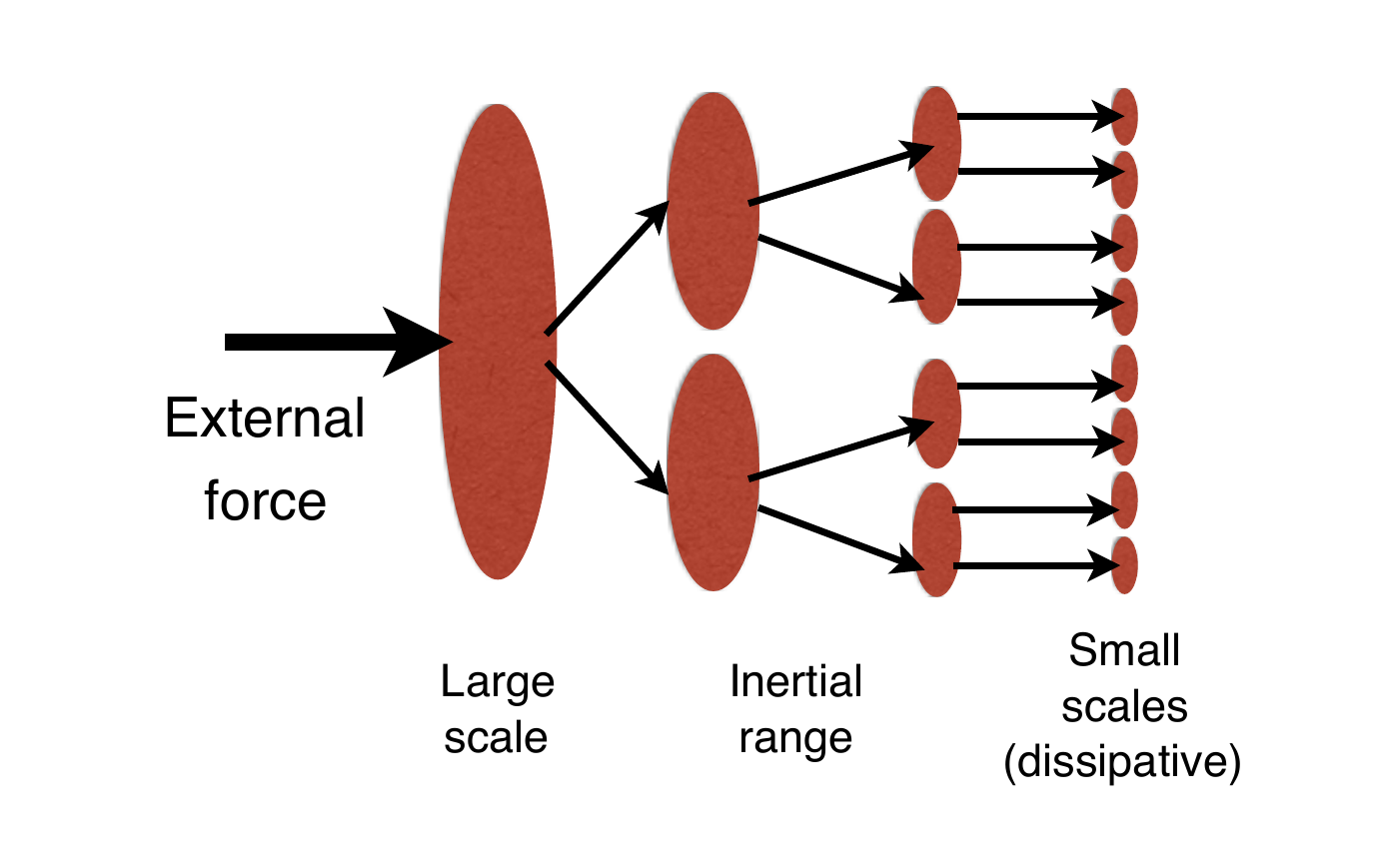}
\caption{Schematic diagrams illustrating energy transfers in three-dimensional hydrodynamic turbulence.  The energy supplied at large scales cascades to the inertial range and then to the dissipative range.  }
\label{fig:NS}
\end{figure}

Can we derive  Kolmogorov's law in the framework of kinetic theory without going to hydrodynamic formulation?  This remains a challenge.  The multiscale flow structures (e.g., vortices within vortices) are natural in the hydrodynamic description of turbulence, but not very apparent in kinetic theory whose basic constituents are particles.     Note however that we can obtain  multiscale fluid structures by    averaging or coarse-graining many times, as is often done in lattice hydrodynamics~\cite{Succi:book:Lattice}.  {\em Yet, the derived structures follow the laws of hydrodynamics, and these laws are not transparent at the particle level. }  Thus, macroscopic description provided by hydrodynamics is much more convenient for  the description of turbulence.  

 Many natural flows involve  more complex forces than those assumed in Kolmogorov's theory of turbulence (see Fig.~\ref{fig:NS}).  For example, Ekman friction, which is of the form $-\alpha {\bf u}$ ($\alpha$ is a positive constant), induces dissipation of kinetic energy at all scales~\cite{Verma:EPL2012}.  Consequently, the energy flux $\Pi_u(k)$ decreases with $k$. Hence, the kinetic energy  in the flow at a given scale is lower than that for $\alpha =0$.  This feature leads to a steeper spectrum for Ekman friction than that predicted by Kolmogorov's theory ($k^{-5/3}$).  Similar steepening of kinetic energy spectrum is observed in buoyancy-driven turbulence~\cite{Obukhov:DANS1959} and in magnetohydrodynamic turbulence~\cite{Verma:ROPP2017}.  A derivation of above variable energy flux is very easy in spectral description of hydrodynamics~\cite{Frisch:book,Lesieur:book:Turbulence,Verma:book:BDF}, but not in kinetic theory.
 
A cautionary remark is in order.  In gas dynamics, kinetic theory is extensively employed to describe rarified gas for which hydrodynamic description breaks down~\cite{Succi:book:Lattice,Singh:PRE2016}.   These ideas find applications in supernova explosions, supersonic rockets and jets, rarified plasma, etc.

%%%%
\section{Dissipation, diffusion, and pressure in hydrodynamics}

In microscopic description of physical processes, the collisions or interactions among  particles conserve  energy.  These processes also respect time reversal symmetry~\cite{Feynman:book:Character,Carroll:book:Time,Pathria:book}.  Given this, it is very difficult to incorporate dissipation for an isolated system of particles.  Hydrodynamic description bypasses this difficulty by postulating viscosity that sets up the energy cascade from  large scales to small scales.  The origin of such friction has been debated by researchers.  In a multiscale hydrodynamic description, the viscosity converts coherent kinetic energy (related to the flow velocity) to incoherent heat energy of microscopic particles at the dissipation scale~\cite{Verma:arxiv:time}; note however that the total kinetic energy at the particle level is conserved.

Turbulence typically enhances diffusion.  We illustrate this phenomena using an often-quoted example---heat diffusion from a heater.  Since  the thermal diffusion coefficient of air is $\kappa \approx 10^{-5}~\mathrm{m}^2/\mathrm{s}$,   from kinetic theory or statistical mechanics, the time estimate for the heat diffusion by $L= 1$~m would be $L^2/\kappa \approx 10^5$ seconds.  This estimate is clearly incorrect.  In reality,  heat is advected by the nonlinear term, hence the time scale is  $L/U \approx 1/0.1 = 10$ seconds, where $U$ is the  velocity of the large-scale structures~\cite{Verma:book:BDF}.  A derivation of the aforementioned hydrodynamic diffusion  from kinetic theory is not practical.   

A related phenomena is {\em Taylor dispersion}~\cite{Taylor:PRSA1954} of particles in a turbulent flow.  The distance between two particles in a turbulent flow increases as $t^{3/2}$, where $t$ is the elapsed time.  Note that the Taylor dispersion is faster than ballistic dispersion ($\sim t$), which is the fastest dispersion for any particle in kinetic theory.  The enhancement in Taylor dispersion  is due to  the advection of the particles by multiscale structures; the particles separated by a distance $r$ hop from  vortices of size $r$ to larger vortices that move with even larger speeds.   Again, Taylor dispersion would be hard to derive in kinetic theory.

As described in  Section~\ref{sec:KT}, the hydrodynamic equations can be derived from kinetic theory.  Such derivations yield equations for compressible flows for which the pressure is the {\em thermodynamic pressure} (that has  origin in kinetic theory).  However, there is another important pressure called {\em dynamic pressure} that appears in incompressible hydrodynamics.  In  Bernoulli's equation, $p+ \rho u^2/2 = \mathrm{constant}$, where $p$ is the dynamic pressure, which is distinct from the thermodynamic pressure. Note that the dynamic pressure   can be derived easily in hydrodynamic framework~\cite{Frisch:book}, but it would be very hard to derive in kinetic theory (without going to coarse-grained picture of hydrodynamics).  We remark that a compressible flow contains both dynamic and thermodynamic pressures~\cite{Zank:PF1991}, but their derivation in kinetic theory would be way too complex.

We conclude in the next section.

\section{Conclusions and Discussions}
In this article, we describe certain hydrodynamic (macroscopic) laws that are difficult to derive {\em directly} from  microscopic framework such as kinetic theory. These laws include Kolmogorov's theory of turbulence,  viscous dissipation and Taylor's dispersion in turbulent flows, and dynamic pressure. For these laws, the hydrodynamic description is more adequate than the kinetic theory.  These observations are in the spirit of discussions by Anderson~\cite{Anderson:Science1972} and Laughlin~\cite{Laughlin:PNAS2000} where they argue in favour of hierarchical description of systems and laws.

We can go to a step (or hierarchy) further in the flow complexity.  Planetary and stellar flows are quite complex; some of the leading problems in  these fields are global warming, ice ages, magnetic field generation, corona heating, mantle and core dynamics of the Earth, land ocean coupling, monsoons, etc.~\cite{Fowler:book}.  To address these problems, particle description is never employed.  Further, it is  impractical (in fact, impossible) to solve the relevant  primitive equations---flow velocity, chemical constituents, moisture,  ice---at all scales.  For the Earth, the corresponding length scales range from $10^{-6}$ m to $4\times 10^6$ m.   Hence, scientists often model these systems using relevant large-scale variables. For example,  ice age is modelled using total solar radiation, carbon dioxide, and the mean temperature of the Earth.  Similarly, the solar magnetic field is modelled using several magnetic modes in spherical harmonic basis~\cite{Jones:book_chapter}.   There are other equally important tools like probability, filtering, and machine learning for describing aforementioned complex systems.  

The next level of hierarchical structures are solar system, galaxy, and the universe.  As we move up the hierarchy, the planetary and stellar atmosphere are ignored and newer sets of variables and equations are used.  For example, Newton assumed the Sun and the Earth to be point particles for describing planetary motion; Millenium simulation of the universe treat the galaxies as point particles embedded in dark matter.

Thus, nature has hierarchical structures that have their own laws and relevant tools~\cite{Laughlin:PNAS2000}.  However,  the system descriptions and associated laws at different levels are connected to each other, most strongly among the neighbouring levels.  For example, kinetic theory and hydrodynamics are intimately connected.  Yet, the laws of the system at a given level are best derived using the equations and tools at that level.   A possible hierarchical  categorisation could be---nuclear and particle physics, atomic and molecular physics, condensed-matter physics, chemistry, biology, ecology, and so on.   Another multiscale characterisation  is---kinetic description of particles, hydrodynamic description of flows, planetary and stellar atmosphere and interiors, solar system, galaxies, galaxy clusters, universe.  These structures help us identify the laws at each level, and derive relationships among them.   It is important to keep in mind that  the connections between the theories at different levels many involve many complications.  Berry~\cite{Berry:PT2002} and Batterman~\cite{Batterman:book:Details} describe such issues, in particular {\em singular limits} encountered in such attempts.

Note that {\em holism}, considered to be the opposite philosophy of {\em reductionism}, advocates that the properties of a system are best understood as a {\em whole}, not as a sum of its parts~\cite{Auyang:book}.  The hierarchical description however differs somewhat from holism, and it propounds that the universe is hierarchically structured, and it is best described by hierarchy of laws at different scales.  These laws however may be interlinked, similar to the laws of kinetic theory and hydrodynamics.   

The  hierarchical framework is often invoked for describing emergent phenomena~\cite{Laughlin:PNAS2000,Laughlin:book,Anderson:Science1972,Anderson:book:More}.   For example, chemists, biologists, and material scientists work tirelessly to discover new molecules and material with specific properties using ab-initio or first-principle calculations.  However, centuries ago researchers used to rely on macroscopic properties of materials (such as, affinity to water, air, fire etc.).  Although no one doubts the power of  first-principle calculations, the former approaches too could be useful.   A major component of climate research involves large-scale simulations of primitive variables on massive grids (say, with one billion grid points).  In comparison, at present, much less attention goes into making a low-dimensional models based on large-scale or macroscopic variables, such as mean temperature, solar radiation, land-sea interactions,  overall carbon dioxide content, etc.  Many believe that a combination of both the approaches, microscopic and macroscopic, would  yield richer dividends.   These illustrations indicate that applications of hierarchical description may help address some of the complex problems we face today.

\section*{Acknowledgments}
I thank Anurag Gupta and Michael Berry for useful discussions.

\end{document}